\newdimen\lineskp
\newdimen\figunit

\documentstyle[12pt]{article}
\newcommand{\mabel}[1]{\label{#1}}
\newcommand{\mibitem}[1]{\bibitem{#1}}

\newcommand{\nn}{\nonumber}

\newcommand{\ignore}[1]{\null}

\renewcommand{\i}{\mm{\rm i}\mm} % originally `is subset of'

\newcommand{\gam}{\gamma}

\newcommand{\lam}{\lambda}
\newcommand{\Lam}{\Lambda}

\newcommand{\e}[1]{\ {\rm e}^{ #1 }\mm}

\newcommand{\spspl}[2]{\frac{\sinh\left(#1\mm+\mm #2\right)}{\sinh\left(#1\mm -\mm #2\right)}\mm} %%%

\renewcommand{\Im}{\mm \mbox{Im}\mm}% was some kinda curly I
\newcommand{\be}{\begin{equation}} %%% be \> {begin{equation}}\\
\newcommand{\ee}{\end{equation}} %%% ee \> {end{equation}}\\
\newcommand{\bea}{\begin{eqnarray}} %%% bea \> {begin{eqnarray}}\\
\newcommand{\eea}{\end{eqnarray}} %%% eea \> {end{eqnarray}}\\
\newcommand{\beastar}{\begin{eqnarray*}} %%% beastar \> {begin{eqnarray*}}\\
\newcommand{\eeastar}{\end{eqnarray*}} %%% eeastar \> {end{eqnarray*}}

\renewcommand{\;}{~ ;~ ~ ~ ~ }
\newcommand{\mm}{\hspace{0.5mm}} %%%
\newcommand{\eq}{Equation }

\newcommand{\half}{{\textstyle \frac{1}{2}}\mm}
\newcommand{\pipergam}{{\textstyle \frac{\pi}{\gamma}}} %%%
\newcommand{\gamperpi}{{\textstyle \frac{\gamma}{\pi}}} %%%

\newcommand{\eps}{\epsilon}
\newcommand{\kap}{\kappa} %%%
\newcommand{\E}{{\cal E}} %%%
\newcommand{\abs}[1]{\left| #1 \right|}
\newcommand{\mat}[2]{\left(\begin{array}{#1} #2 \end{array}\right)}
\newcommand{\longat}[1]{\hskip 0.5mm {\rule[-4mm]{.15mm}{8mm}}_{\hskip 0.5mm #1}} %%%

\newcommand{\mod}{\ \mbox{mod}\ }
\newcommand{\unit}{1\!\!\hskip 1pt\mbox{l}}

\def\sZZ{{\rm Z\kern-0.32em{}Z}}
\def\ZZ{{\mathchoice
  {\sZZ}
  {\sZZ}
  {\rm Z\kern-0.30em{}Z}
  {\rm Z\kern-0.25em{}Z} }}

\newcommand{\NPB}{{~Nucl. Phys.} B~}
\newcommand{\PRL}{{~Phys. Rev. Lett. }}

\newcommand{\PRB}{{~Phys. Rev.} B~}

\newcommand{\PLB}{{~Phys. Lett.} B~}

\newcommand{\JPA}{{~J. Phys.} A~}

\newcommand{\RMP}{{~Rev. Mod. Phys. }}
\begin{document}
\baselineskip= \lineskp plus 1pt minus 1pt

\ignore{}
%{{{ Titlepage

%\begin{titlepage}
\pagestyle{empty}

\rightline{NORDITA-97/33 P}
\rightline{hep-th/9707098}
\rightline{   }
\rightline{July 10, 1997}

\vskip 0.5truein
\begin{center}

{\Large\bf Excited States in the Twisted XXZ Spin Chain
}\\
\vskip 0.5truein
{\bf T. Nassar$^a$$^*$ and O. Tirkkonen$^b$${}^{\dag}$
}\\
\medskip
{\it $(a)~$Laboratoire de Physique Math\'ematique - Universit\'e de
Montpellier II \\ CNRS-URA 768 - 34095 Montpellier Cedex 05,
France\\and ~~
\\ Helsinki Institute for Physics, P.O. Box 9, FIN-00014 University
of Helsinki, Finland\\~~
\\$(b)~$NORDITA, Blegdamsvej 17, DK-2100 Copenhagen, Denmark}\\

\bigskip
\vfill
%\vskip 1.5 truein

{\bf Abstract}

\end{center}

%\begin{abstract}

\noindent
We compute the finite size spectrum for the spin 1/2 XXZ chain with
twisted boundary conditions, for anisotropy in the regime
$0<\gam<\pi/2$, and arbitrary twist $\theta$. The string hypothesis is
employed for treating complex excitations. The Bethe Ansatz equtions
are solved within a coupled non-linear integral equation approach,
with one equation for each type of string. The root-of-unity quantum
group invariant periodic chain reduces to the XXZ${}_{1/2}$ chain with
a set of twist boundary conditions ($\pi/\gam\in\ZZ$, $\theta$ an
integer multiple of $\gam$). For this model, the restricted Hilbert
space corresponds to an unitary conformal field theory, and we recover
all primary states in the Ka\v c table in terms of states with
specific twist and strings.

\vskip 15mm
%\centerline{PACS: }

\begin{flushleft}
\rule{5.1 in}{.007 in}\\
$^{*}$ {\small E-mail: nassar@rock.helsinki.fi \\ }
$^{\dag}$ {\small E-mail: tirkkone@nordita.dk \\ }
\end{flushleft}

%\end{abstract}
%\end{titlepage}
\pagebreak
\pagestyle{plain}
%\clearpage
%\pagenumbering{roman}
%\setcounter{page}{2}
%\tableofcontents
%\pagenumbering{arabic}
\setcounter{page}{1}   %after titlepage or \section or pgbrk -> no fuss

\newpage

%}}}

%{{{ Introduction

\section{Introduction}

Apart from their relevance to real lattice systems with restricted
dimensionality, spin chains are interesting from a field
theoretical point of view as integrable lattice regularizations of
1+1 dimensional field theories \cite{kogut}.

The continuum limit of the field theory is the
thermodynamic limit of the spin chain, where the number of
spins in the chain goes to infinity. If the thermodynamic
limit is accompanied by taking the lattice spacing to zero,
keeping the length of the chain constant, we get a
description of the continuum field theory in finite volume.
At criticality, one expects conformal invariance, with the
conformal weights read out via the finite size corrections
of the energy and the momentum
\cite{blocani,affleck,cardy}.

In the class of theories with trigonometric $R$-matrices, massive
theories (sine-Gordon/massive Thirring models) are usually associated
to spin-1/2 anisotropic Heissenberg chains (XXZ${}_{1/2}$ chains) with
inhomogenieties \cite{izekor}, whereas massless theories are related to
homogeneous chains. The conformal properties of the latter are well
known, the central charge $c=1$, and conformal weights are ones of the
gaussian model \cite{devekar,woyeck,boizre}.

Central charges $c<1$ can be reached by considering chains with some
special fixed, or twisted boundary conditions, related to critical
Potts and Ashkin-Teller chains
\cite{devekar,alibaba,haquibat,albagrri}. The correct interpretation
of this lowering of $c$ in the case of open chains was found by
Pasquier and Saleur \cite{passal}. The key feature is the invariance
of the spin chain dynamics under the universal quantum enveloping
algebra $U_q(sl_2)$. (The quantum deformation parameter is given by
the spin chain anisotropy as $q=\e{\i\gam}$). For open chains, the
boundary conditions can be chosen to ensure $U_q(sl_2)$ invariance.
This, however, is not sufficient to alter the central charge, which is
a bulk effect. The lowering of $c$ comes about for a root-of-unity $q$
(rational $\gam/\pi$), when the quantum group invariance allows for a
self-consistent truncation of the Hilbert space. By restricting the
Hilbert space to type-II (a.k.a ``good'') representations of the
quantum group, the central charge is lowered from $c=1$ to those of
minimal models \cite{passal}. Furthermore, analyzing low lying excited
states shows that the conformal weights of primary operators form a
one-parameter subset of the
 Ka\v c table.

For closed chains, quantum group invariance is a more delicate matter
\cite{karzap,grpaprra}. A topological interaction of the
Wess-Zumino-Witten type has to be introduced \cite{karzap}, which
leads to non-local terms in the Hamiltonian. On the level of Bethe
Ansatz equations, the quantum group invariant closed chain can be
viewed as a collection of XXZ${}_{1/2}$ chains with a set of twisted
boundary conditions commensurate with the anisotropy. Again, quantum
group reduction leads to a lowering of the central charge to $c<1$
\cite{grpaprra,papr}. The collection of twist boundary conditions is
the exact analogue to the $\theta$-vacua of SOS and sine-Gordon
models, in which the quantum group restriction leads to RSOS or RSG
models, respectively. The quantum group invariant chain can be
considered as the critical/ultraviolet limits of these models (see
e.g. Ref. \cite{fimaqura}).

The main motivation for this work is the interpretation
found in Reference \cite{fadtir}, of the conformal
properties of twisted XXZ${}_{1/2}$ chains, in terms of a
lagrangean conformally invariant field theory with $c<1$,
namely the Liouville model with imaginary coupling. If the
imaginary Liouville model on a circle is discretized
preserving the integrable structure, it is equivalent to
the quantum group invariant periodic XXZ${}_{1/2}$ chain on
the level of Bethe Ansatz; the Liouville model can be
mapped to a set of XXZ${}_{1/2}$ chains with twisted
boundary conditions. For root of unity couplings, a quantum
group reduction can be performed on the Hilbert space. The
central charges of minimal models are reproduced,
confirming the dependence of central charge on Liouville
coupling, known from canonical quantization.

Furthermore, based on Karowski's treatment of Potts models
\cite{karowski}, primary states furnishing the full
 Ka\v c tables of unitary minimal models were conjectured to arise
from a specific set of Bethe Ansatz states. More exactly, primary
states should arise from BA states with one string excitation above
the sea of real BA rapidities, and no extra holes, i.e. vanishing
total spin $S$ of the chain.

This leads us to reconsider the finite size analysis of
XXZ${}_{1/2}$ chains with twisted boundary conditions. In the
extensive literature on the subject
 \cite{devewoy,woyeck,boizre,alibaba,haquibat,karowski,woectr,%
eckham,kluwezi,suthshast,yufow,roecksuth,papr},
 the main stress has been on investigations of pure hole
excitations. Very little has been done to analyze complex
excitations. String excitations with a restricted set of couplings
and twists were investigated in \cite{karowski}, and complex pair
excitations were considered in \cite{kluwezi}. Apart from these,
most work on complex excitations has been numerical
\cite{alibaba,yufow,roecksuth}.

We shall use a non-linear integral equation (NLIE) method
\cite{peaklu,desdeve} to analyze the finite size spectrum. This
method is more powerful than the Euler-MacLaurin methods mostly
used in the literature, and it allows for extracting the scaling
behaviour from the Bethe Ansatz equations in a rather straight
forward manner. We follow the Destri-de Vega (DdV) approach
\cite{desdeve}, extending it to complex excitations with a twist.
In Reference \cite{ddv2}, the DdV equations for the sine-Gordon
model were generalized to complex excitations described as wide and
close pairs. Here, we make the crucial departure from the methods
of \cite{ddv2} in that the string picture of the excitations is
used. This is because, as pointed out above, the primary states are
most transparently described in this language.

The plan of the paper will be the following: In Section 2, we
formulate the albegraic Bethe Ansatz for the XXZ${}_{1/2}$ chain with
twisted boundary conditions. This we do for generic anisotropy in the
antiferromagnetic regime $0< \gam < \pi/2$ and generic twist angle
$\theta$. We describe the complex excitations as strings. The DdV
equations are derived for the case where we have positive parity
strings of one specified length $k$ in addition to a bulk of real
Bethe Ansatz roots. To facilitate the analysis, we restrict the range
of string lengths to obey $k \gam< \pi$. In Section 3, we compute the
finite size corrections to the energies and momenta of these states.
We specialize to states where no holes punctuate the distribution of
real roots. In Section 4, we specialize to the Hilbert space
pertinent for lattice Liouville model, i.e. the restricted quantum
group invariant chain. We show how the quantum group reduction lowers
the central charge. Analyzing excited states with only one longer
string in the sector $S^z=0$, we show that these states are primary
states of the theory, furnishing the whole
 Ka\v c table.

The main point of this paper will be the fact that taking Bethe
Ansatz configurations with one positive parity string, we are able
to retrieve the spectrum of unitary minimal conformal field
theories using a NLIE method. The two integers in the ensuing
 Ka\v c table are the string length $k$ and the twist $\kappa$.
Aside from the technical advantage gained from this way of doing
things, the hope is that this method is easier to  generalize
to the less explored case of $c>25$ of relevance to Liouville
theory with real coupling.

%}}}
%{{{ String BAE

\section{Twisted Bethe Ansatz equations for strings}

We consider a spin 1/2 XXZ chain with $2N$ sites, with anisotropy
$\gam$, and boundary conditions twisted by the angle $\theta$:
\be
 \sigma_{2N+1}^\pm = \e{\pm\i\theta}  \sigma_{1}^\pm \;
 \sigma_{2N+1}^z =  \sigma_{1}^z
~.
\ee
 The corresponding hamiltonian reads:
\be
 H_{XXZ}=\half \sum _{j=1}^{2 N} \sigma _{j}^{+}\sigma _{j+1}^{-}
        + \sigma _{j}^{-}\sigma _{j+1}^{+}
        + 2 \cos \gam \sigma _{j}^{z}\sigma _{j+1}^{z}
\ee
 where the antiferromagnetic XXX limit corresponds to $\gam =0$
and the free Fermion point is at $\gam =\frac{\pi }{2}$.

We formulate the theory in the standard language of quantum inverse
scattering \cite{koboiz}. The L-matrix, satisfying the fundamental
commutation relation $RLL=LLR$, is
\be
 L_{n,a}(\lam) = \frac{1}{\sinh \i\gam}\mat{ll}{
 \sinh(\lam + \i\gam\mm \sigma_{n}^z) &
		\sigma_{n}^- \mm \sinh\i\gam\\
 \null&\null\\
 \sigma_{n}^+ \mm \sinh\i\gam &
		\sinh(\lam - \i\gam\mm \sigma_{n}^z) } ~ .
 \mabel{Lxxz}
\ee
 where the operators $\sigma_{n}^{z},\sigma_{n}^{\pm }$ are defined by
the commutation relations of the quantum group $U_{q}(sl_{2})$. The
L-matrix is normalized so that for $\lam = \i\gam/2$ it degenerates to
a permutation operator between the auxiliary and quantum spaces. Using
this L-matrix, we can write the transfer matrix of the chain, and
build the eigenstates of the transfer matrix using the algebraic Bethe
Ansatz.

We will focus our attention on those eigenstates of the transfer
matrix that can be described in terms of real roots $\lam_j,~ j=1,
... n_1$, as well as strings of complex roots $\lam^{(l)}_{j},~
j=1, \ldots ,n_l$. A (positive parity) $l$-string $\lam^{(l)}$ is
the collection of $l$ roots of the form $\lam + \i n \gam~,
~n=-(l-1)/2, ~\ldots,~ (l-1)/2$, with $\lam$ real. Accordingly, in
compact notations real roots are frequently designated as
one-strings.

We shall only consider positive parity strings. Negative parity
strings can be treated within the same context but they are
irrelevant for our purposes since they do not lead to primary
conformal states in the thermodynamic limit.

The number of $l$-strings is $n_l$, and the total number of roots
is $\sum_l ~l~ n_l = N - S$, where $S$ is the number of Bethe roots
lacking from the maximal number of roots $N$, i.e. the
$z$-component of the total spin of the chain.
 In order to simplify the underlying analysis, we consider only the
antiferromagnetic regime $\gamma<\pi/2$, which covers all unitary
theories. We also adopt a technical restriction on the range of the
string length, so that $k \gamma < \pi$. This restriction, although
not required in the most general case of anisotropy, shows up
rather naturally when we consider the roots of unity case of the
XXZ model pertaining to unitary conformal field theories, as a
consequence of the underlying quantum group structure.

The Bethe Ansatz equations solved by the rapidities
$\lam^{(k)}_{i}$ describing a Bethe state are
\be
 \e{-2\i k\theta}\left[\spspl{\lam^{(k)}_{i}}{\i k \gam/2} \right]^{2N}
= -\prod_{l, ~n_l\neq 0} ~ \prod_{j=1}^{n_l} S_{kl}(\lam^{(k)}_{i} -
\lam^{(l)}_{j})~,~ ~ ~
\forall (k,i) ~, \mabel{BAE}
\ee
 where the string-string scattering matrix is
\be
 S_{kl}(\lam) = \prod_{m=\abs{k-l}/2}^{(k+l)/2 -1}
 \spspl{\lam}{\i m\gam} \spspl{\lam}{\i (m+1) \gam} ~.
\ee

To solve these equations, it is beneficial to write them in a
logarithmic form. For this, we define the  phase function
\bea
 \phi_m(\lam) &=& \i\log\spspl{\i m\gam}{\lam},~ m>0 \mabel{phim} \\
 \phi_0(\lam) &=& 0 ~. \nn
\eea
For real argument $\lam$, the phase function $\phi_m$ is a continuous
monotonic function. We choose the branch of the logarithm so that it
is antisymmetric with respect to the origin. Using the phase function,
we further define the momentum function
\be
 p_l(\lam) = \phi_{l/2}(\lam) ~,
\ee
  the string-string scattering phase
\be
 \Phi_{kl}(\lam) =
\sum_{m=\abs{k-l}/2}^{(k+l)/2-1} \bigl[ \phi_m(\lam)+\phi_{m+1}(\lam) \bigr]~,
\ee
 and  the counting functions for $l$-strings
\be
 z_l(\lam) = 2N p_l(\lam) - \sum_{k,j} \Phi_{lk}(\lam - \lam^{(k)}_{j})
 ~+~ 2 l\theta ~. \mabel{zk}
\ee
  By defining counting functions for strings, all the
analysis is reduced to dealing with functions of real
variables, a considerable simplification over the usual
method of treating complex NLIE's. In this respect, the
string hypothesis is necessary, (although probably not
inescapable), and marks a departure from the standard
Destri-de Vega treatement. Of course, we must keep in mind
that this is only justified when we are at zero
temperature.

In terms of the counting functions, the Bethe equations (\ref{BAE})
become
 \be
 z_l(\lam^{(l)}_{i}) = 2\pi I^{(l)}_{i} ~, \mabel{IBAE}
\ee
 where the quantum numbers $I^{(l)}_{i}$ encode the choice of branch
of the logarithm. They are integers or half integers when $n_l$ is
even or odd, respectively. These quantum numbers should all be
distinct for a specified string length, otherwise the Bethe Ansatz
equations are easily seen to lead to a infinite repulsion of the
quasiparticles corresponding to equal quantum numbers. This is what is
usually referred to as the fermionic character of Bethe states.

The basic assumption one makes in treating the Bethe equations for
real rapidities $\lam$ is that the monotonicity of $2 N p_1$ is
sufficent to make the counting function $z_1$ overall monotonic. This
allows one to find the spectrum of the quantum numbers $I^{(1)}_{i}$.
Indeed, assuming monotonicity, $z_1/2\pi$ takes all the (half)
integer values between $z_1(-\infty)/2\pi$ and $z_1(\infty)/2\pi$
exactly once. The only freedom we are left with in determining the
spectrum of the $I$:s lies in the choice of the overall branch of the
counting function $z_1$, which is irrelevant. Any possible
non-monotonicity would induce additional holes close to the ends of
the rapidity distributions \cite{ddv2}. These don't effect the
quantities calculated here. As opposed to $z_1$, the higher string
counting functions $z_k$ are generically non-monotonous.

For a configuration characterized by the higher string occupation
numbers $n_k$, the asymptotic values of the counting functions are
\be
  z_l(\pm\infty) = \pm \pi n_l ~\pm~ 2 S (\pi-l \gam)
           ~\pm~ 2\pi\sum_{k>l} (k-l)n_k ~+~ 2 l\theta ~. \mabel{asymz}
\ee
 If $n_k \neq 0$ for some $k$, it is clear that there are more
vacancies between $z_1(-\infty)$ and $z_1(\infty)$ than those
occupied by the Bethe Ansatz roots. These unoccupied vacancies are
called holes, and they correspond to zeros of the eigenvalue of the
transfer matrix (as opposed to Bethe Ansatz roots, which correspond
to poles with vanishing residue). The number of holes in the
distribution of real roots is
\be
 h = 2\sum_{k>1} (k-1)n_k ~+~ 2 S
~+~ \left\lfloor \frac12 - \frac\theta\pi
                       - S \frac\gamma\pi \right\rfloor
~-~ \left\lfloor \frac12 - \frac\theta\pi
                       + S \frac\gamma\pi \right\rfloor
  ~. \mabel{holes}
\ee
 Here $\lfloor \circ \rfloor$ denotes the integer part of $\circ$.
The integer part terms vanish for small values of the total spin
$S$. The salient feature of Relation (\ref{holes}) is that a
$k$-string gives rise to $2(k-1)$ holes.

The antiferromagnetic ground state is given by the configuration
where there are no complex roots, and the number of real roots is
maximal, i.e. $n_1=N;~ n_k=0,~ k>1;~ S=0$. From \eq (\ref{holes}),
it is easily seen that there are no holes; in the ground state all
vacancies are filled.

In the thermodynamical limit, low-lying exitations above the ground
state are characterized by configurations where some higher strings
are exited, as well as possibly some extra holes ($S$ might differ
from zero), but the only macroscopic occupation number is the one of
real roots, i.e. only $n_1 \sim N$.

Now we make use of the Destri - de Vega method \cite{desdeve} to
derive an integral equation equivalent to the Bethe equations. The
derivative of the function
\be
 A(\lam) =  \log\{1+(-1)^{n_1}\e{\i Z_1(\lam)}\}
\ee
 acts as a density of real roots and holes in contour integrals; it
has first order poles at all allowed $2\pi~\times$ (half) integer
values of $z_1(\lam)$, with residue one. Accordingly, for a
function $f$ analytic within the contour,
\be
 \oint_\Gamma f(\lam)~ A' = \sum_{j:{\rm roots}} f(\lam_{j})
 ~+~ \sum_{m:{\rm holes}} f(\mu_{m}) ~. \mabel{ddvtrick}
\ee
 The integration path $\Gamma$ surrounds the roots and holes
counterclockwise. Stretching $\Gamma$ to $\pm\infty\pm\i\eps$, one
gets the coupled equations for the counting functions of real roots
and $k$-strings
\be
 z_1(\lam) = 2N\sigma(\lam)
 ~+~ 2\Im \Bigl(G_{11} *_\epsilon A\Bigr)(\lam)  %\cr &&
 ~+~ \sum_{m=1}^h F_{11}(\lam-\mu_m)
 ~-~ \sum_{j=1}^{n_k} F_{1k}(\lam-\zeta_j)
 ~+~ C_1 \\
\ee
\be
z_k(\lam) = 2\Im \Bigl(G_{1k} *_\epsilon A\Bigr)(\lam) %\cr &&
 ~+~ \sum_{m=1}^h F_{1k}(\lam-\mu_m)
 ~-~ \sum_{j=1}^{n_k} F_{kk}(\lam-\zeta_j)
 ~+~ C_k ~.
\ee
 These equations generalize the results of Ref. \cite{desdeve} to
encompass strings and holes. \footnote{The generalization to pure
hole excitations was performed in Ref. \cite{fimaqura}.} We have
denoted the rapidities of the holes with $\mu_m, ~m=1,\ldots,h$, and
the rapidities of the $k$-strings with $\zeta_j, ~j=1,\ldots n_k$.
The symbol $*_\epsilon$ denotes a convolution along a infinitesimally
shifted real line,
 \be
(G*_\epsilon A)(\lam)
 \equiv \int\frac{d\mu}{2\pi} G(\lam - \mu - \i\eps) ~A(\mu + \i\eps)
 ~.
 \mabel{epsconv}
 \ee

The inhomogeneity function $\sigma$, arising from the action of an
inverse convolution on the momentum function, is
\be
 \sigma(\lam) = (\unit + \Phi_{11}')^{-1}*p_{1}
 = \arctan\sinh \pipergam  \lam ~. \mabel{sigma}
\ee
 Similarly, the functions $F$, giving the hole and string contributions,
are defined as
 \be
 F_{1l} = (\unit + \Phi_{11}')^{-1}*\Phi_{1l}  ~ .
 \ee
  The measures of the convolutions have  $2\pi$ denumerators, as in
(\ref{epsconv}), and $\unit(\lam) \equiv 2\pi\delta(\lam)$. The
kernels $G$ of the integral equations are given by differentiation,
 $
 G_{1l}(\lam) = \frac{d}{d\lam} F_{1l}(\lam)
 $,
 and the effect of the $k$-string on itself is encoded by the function
 $
 F_{kk} =  \Phi_{kk}  - G_{1k} * \Phi_{1k}
 $.

It is remarkable that the $z_k$ equation doesn't have any
inhomogeneity arising from the momentum functions. This follows from
the identity $p_k = G_{1k} * p_1$, which can be proved by Fourier
transforming.

Finally, the constant terms in the equations are
\be
 C_1 = \frac{\pi~ \theta}{\pi-\gam}  -
 \frac{\pi(\pi-2\gam)}{2(\pi-\gam)}
 \left( \left\lfloor  \frac12 - \frac{\theta}{\pi}
                     + S\frac\gamma\pi \right\rfloor  +
        \left\lfloor  \frac12 - \frac{\theta}{\pi}
                   -S\frac\gamma\pi \right\rfloor \right)
~ ~ ~ ~ ~ ~ ~ ~ ~ ~ \ee \be
 C_k = \frac{(2k-2)\pi~ \theta}{\pi-\gam}  -
 \frac{\pi(\pi-k\gam)}{\pi-\gam}
 \left( \left\lfloor  \frac12 - \frac{\theta}{\pi}
                     + S\frac\gamma\pi \right\rfloor  ~+~
        \left\lfloor  \frac12 - \frac{\theta}{\pi}
                   -S\frac\gamma\pi \right\rfloor \right)
\ee
 These have the contributions of of the twist angle $\theta$,
renormalized by the integration over the inverse convolution $(\unit +
\Phi_{11}')^{-1}$. The integer part terms arise from boundary terms in
partial integrations.

\bigskip

For computing the finite size corrections, one takes $N$ large but
not infinite. From the functional form of the source term,
$2N\sigma$, one can see that the Fermi points, i.e. the limits of
the distribution of real rapidities, are close to $\lam=\pm\log N$.
To treat the excitations close to the Fermi points, one defines the
so called kink counting functions,
\bea
  z_1^\pm(\lam) &=& \lim_{N\to\infty} \Bigl(z_1(\lam \pm \gamperpi \log 4N)
~\mp~ N \pi \Bigr) \\
  z_k^\pm(\lam) &=& \lim_{N\to\infty} z_k(\lam \pm \gamperpi \log 4N)
  ~.
\eea
 The configurations of holes and strings that can be treated
exactly in this limit are such that all holes and strings are
either close to one of the Fermi points, or well away from both.
Here, we shall assume that every hole and string is in the vicinity
of one of the Fermi points.

We parametrize the hole rapidities as
 \be
 \mu_m = \mu^\pm_m \pm \gamperpi \log 4N ~, ~ ~ m=1,\ldots,h^\pm ~.
\mabel{holerap}
\ee
 The number of holes in the vicinity of the left (right) Fermi
point is $h^+$ ($h^-$). The total number of holes is $h=h^+ + h^-$.
Similarly, the $k$-string rapidities are $\zeta_j = \zeta_j^\pm \pm
\gamperpi \log 4N$, and $n_k^\pm$ are the numbers of strings close
to each Fermi point. At one Fermi point, the contributions of the
holes and strings situated close to the opposite Fermi point are
just given by $F(\pm\infty)$.

The only non-trivial scaling behaviour comes from the function
$\sigma$. Indeed, subtracting the bulk contribution, one has for values of $\lam$ close to the Fermi points
\be
 \lim_{N\to\infty} \Bigl( 2 N \sigma(\lam\pm \gamperpi \log 4N) \mp N
\pi \Bigr) = \mp \e{\mp\pipergam  \lam} ~.
\ee

We get the following coupled non-linear integral equation for the
kink counting functions:
\be
  z_l^\pm(\lam) = \Omega_l(\lam)
     ~+~ 2 \Im\Bigl( G_{1l}*_\eps A_\pm \Bigr)(\lam) ~,~ ~ l = 1,k ~,
 \mabel{kink}
\ee
 where we have used the  scaled density integral
 $
A_\pm = \log\Bigl(1+(-1)^{n_1}\e{\i z^\pm}\Bigr)
 $, and the inhomogeneity functions
\be
 \Omega_l(\lam) =
 \mp ~\delta_{1,l} ~\e{-\pipergam  \lam}
 ~+~ \sum_{m=1}^{h^\pm} F_{1l}(\lam-\mu^\pm_m)
 ~-~ \sum_{j=1}^{n_k^\pm} F_{lk}(\lam-\zeta_j^\pm)
 ~+~ C_l^\pm  ~.
 \mabel{inhomo}
\ee
  The constant terms carry the renormalized twist $\theta$, as well
as the information from the opposite Fermi point:
\be
 C_l^\pm = C_l ~+~ h^\mp F_{1l}(\pm\infty)
               ~-~ n_k^\mp F_{lk}(\pm\infty)
  \mabel{constantspm}
\ee

For extracting energy and momentum eigenvalues from the scaled
NLIEs (\ref{kink}), one needs to know the asymptotic behaviour of
the scaled functions. At $\pm\infty$ we have
 \be
 z_1^\pm(\pm\infty) ~=~ \pm \pi n_k (k-2) \pm \pi S \mp 2 S\gam +
2\theta ~ .
 \ee
  At $\mp\infty$, the kink equation (\ref{kink}) implies
 \be
 z_1^\pm(\mp\infty) = \mp z_1^\pm{}'(\mp\infty) = \mp\infty ~.
 \ee

Due to the infinitesimal $\eps$ in $\Im A(\lam+\i\eps)\equiv \Im_\eps
A $, the value of this expression always between $-\pi/2$ and
$\pi/2$. In the limit $\eps\to 0$, it is a sawtooth function, with a
step-function jump of $-\pi$ at each root or hole position. A small
finite $\eps$ makes the function analytic. The asymptotic values of
the imaginary parts of the scaled density integrals are thus
\be
 \Im_\eps A_\pm(\pm\infty) =
 \Im_\eps A(\pm\infty) ~=~  \theta ~\mp~S \gam
 + \pi \left\lfloor  \frac12 - \frac{\theta}{\pi}
   \pm S\frac\gamma\pi \right\rfloor ~\equiv~ \theta_\pm ~\mp~ S \gam
 \mabel{asymA+} ~,
\ee
 It should be noticed that close to $\pm\infty$ the imaginary part
operation can glide through the $G_{1l}$ to act only on $A$, up to
a correction that vanishes like $\epsilon\log\epsilon$. That is,
the integrals in the convolutions of Equation (\ref{kink}) are taken
along the real axis, so that the kernels $G_{1l}$ are convoluted
with the regularized sawtooth function $\Im_\eps A$.

In the opposite asymptotic regime, deep in the bulk of filled real
rapidity states, the small but finite $\eps$ lets
$z_1^\pm{}'(\mp\infty)$ dominate in the expansion of
$A_\pm(\mp\infty+\i \eps)$, and
\be
 \Im_\eps A_\pm(\mp\infty) = 0  ~. \mabel{asymA-}
\ee

%}}}
%{{{ The finite size spectrum

\section{The Finite Size Spectrum}

At the external rapidity $\lam$ close to $\i\gam/2$, the dominating
contribution to the eigenvalue of the transfer matrix is given by
 \be
\Lam(\lam;\{\lam^{(k)}_{i}\}) = \e{-\i\theta}
\left[\frac{\sinh(\lam+\i\gam/2)}{\sinh{\i\gam}}\right]^{2N}
\prod_{l,~n_l\neq 0} ~ \prod_{j=1}^{n_l}
\frac{\sinh(\lam^{(l)}_{j}-\lam+\i(l+1)\gam/2)}{\sinh(\lam^{(l)}_{j}-\lam-\i(l-1)\gam/2)}~.
 \ee
 We expect local integrals of motion at the point $\lam =
\i\gam/2$, where the $L$-matrix (\ref{Lxxz}) degenerates to a
permutation operator. The corresponding momentum and energy
eigenvalues of a Bethe state are
\be
 P(\{\lam^{(l)}_{j}\}) =  \i\log\Lam(\i\gam/2;\{\lam^{(l)}_{j}\})
 ~=~ \Bigl( \pi \sum n_l + \theta + \sum_{l,j} p_l(\lam^{(l)}_{j}) \Bigr)
~{\rm mod}~ 2
\ee
\be
 E(\{\lam^{(l)}_{j}\}) =  \i\frac{\gam}{\pi}\mm \frac{d}{d\lam}\mm
 \log\Lam(\lam;\{\lam^{(l)}_{j}\}) \longat{\lam=\i\frac{\gam}{2}}
 \!\!\!\!~=~ 2 N \frac\gam\pi \cot\gam - \frac\gam\pi \sum_{l,j}
  p_l'(\lam^{(l)}_{j}) ~.~ ~ \mabel{ene-ene}
\ee
 Using the logarithmic Bethe equations (\ref{IBAE}), we express the
momentum in terms of the quantum numbers $I$:
\be
P(\{\lam^{(l)}_{j}\}) = \Bigl( \pi \sum n_l ~+~ \frac\pi {N} \sum_{l,j}
I^{(l)}_{j} \Bigr) ~{\rm mod}~ 2
  ~. \mabel{moment-mal}
\ee

To compute the energy eigenvalues, we once more employ the contour
integral method (\ref{ddvtrick}) for the sum over rapidities of
real roots $\lam^{(1)}_{j}$. Doing this, we get
\bea
 E(\{\mu_{m}\};\{\zeta_j\}) &=& 2N \frac\gam\pi \left( \cot\gam -
\int_{-\infty}^\infty \frac{d\lam}{2\pi} p_1'(\lam) \sigma'(\lam)
\right)
  \cr & & \cr & &
 ~+~ \frac\gam\pi\Im_\eps\int_{-\infty}^\infty\frac{d\lam}\pi
\sigma'(\lam)A'(\lam)
 ~+~ \frac\gam\pi \sum_m \sigma'(\mu_m) \mabel{energy-mal}
\eea
 The dependence of the string rapidities $\zeta_j$ is only
indirect, coming through the Bethe equations in the NLIE form, when
evaluating the last two terms.

The $1/N$ corrections to the energy come from the parts of the
second integral in (\ref{energy-mal}) that are close to the Fermi
points $\lam = \pm \gamperpi \log 4 N$, as well as from the third
term. From \eq\ (\ref{sigma}) we have
 $
 \sigma'(\lam) = \frac\pi\gam\mm (2\cosh \frac\pi\gam \lam)^{-1}
 $,
 which is exponentially peaked around the origin. From the
asymptotic analysis of $\Im_\eps A$, we however know that for large
$N$, $\Im_\eps A'$ vanishes as a double exponential when one goes
in to the bulk from the Fermi points $\pm \gamperpi \log 4 N$. Thus
the main contributions to the second integral in (\ref{energy-mal})
for large $N$ come from the vicinity of these points. In this
regime,  $\sigma'$ takes an exponential form,
 $
  \sigma'(\mu \pm \gamperpi \log 4 N)
 \approx \frac\pi\gamma \frac1{4N} \e{\mp \pipergam \mu}
 $.
 Also, the hole rapidities in the last term of (\ref{energy-mal})
are taken to be close to one of the Fermi points, as in \eq\
(\ref{holerap}).

After these considerations, the finite size corrections to the
energy are given by
\be
 E_{{\rm FSC}} = \frac 1N ~(\E_+ + \E_-) ~,
\ee
 where the contributions of the right and left ends of the rapidity
spectrum are
\be
 \E_\pm =  \Im_\eps\int \frac{d\mu}\pi \e{\mp \pipergam \mu} A_\pm'
  ~+~ \sum_{m=1}^{h^\pm} \e{\mp\pipergam\mu_m^\pm} ~.
\ee

Using the scaled NLIEs (\ref{kink}), these can further be
reduced to
\be
 \E_\pm = \mp \sum_{m=1}^{h^\pm} z_1^\pm(\mu_m^\pm)
 ~\pm~ \sum_{j=1}^{n_k^\pm}  z_k^\pm(\zeta_j^\pm)
 ~\pm~ h^\pm C_1^\pm  ~\mp~ n_k^\pm C_k^\pm
 ~\pm~ \Im_\eps \int \frac{d\mu}\pi ~\Omega_1^\pm{}'(\mu)
A_\pm(\mu) ~,
 \mabel{calE}
\ee
 where the $\Omega$:s are defined in \eq\ (\ref{inhomo}).

The terms in (\ref{calE}) involving the constants $C_1$ and $C_k$
of Equation (\ref{constantspm}) can be evaluated using the
asymptotic calues of the $F$-functions: $F_{11}(\pm\infty) = \pm\pi
\mp \frac 12 \frac{\pi^2}{\pi-\gam}$ and $F_{1k}(\pm\infty) = \pm k
\pi \mp (k-1) \frac{\pi^2}{\pi-\gam}$.

The values of the counting functions at holes and roots are given by
$2\pi$ times the (half)integer quantum numbers $I^\pm$. To evaluate
the sums over the hole quantum numbers in (\ref{calE}), we take a
distribution where all real roots are in the bulk, so that no holes
intermeddle with the roots (Fig 1.). Thus the quantum numbers
$I^{+}_{m}$ for the positive rapidity holes get the $h^+$ largest
(half)integer values of $z_1^+(\lam)/2\pi$, and the quantum numbers
$I^{-}_{m}$ for the negative holes get the $h^-$ smallest
(half)integer values of $z_1^-(\lam)/2\pi$:
 \be
I^{\pm}_{m} ~=~ \pm \frac12 \Bigl( (k-2)n_k + S - 1 \Bigr)
~-~\left\lfloor\frac12-\frac\theta\pi \pm S \frac\gam\pi\right\rfloor
            ~\mp~ 1 ~\pm~ m,~ ~ m = 1, \ldots h^\pm
 ~. \nonumber
 \ee
 The quantum numbers $I^{k,\pm}$ of the strings may take any
(half)integer values allowed by the asymptotics of the counting
functions (\ref{asymz}) up to non-monotonicities that are consistent
with the non-linear integral equations (\ref{kink}).

Finally, the integral in the last term in (\ref{calE}) can be
evaluated by the standard dilogarithm trick, see e.g.
\cite{desdeve,ddv2}. For the densities $A_\pm$ and the inhomogenities
$\Omega^\pm$ satisfying equations of the form (\ref{kink}), and the
boundary conditions (\ref{asymA+},\ref{asymA-}) for $A_\pm$, we have
\be
  \Im_\eps\int_{-\infty}^\infty d\mu~ \Omega^\pm{}'(\mu) A_\pm(\mu)
~=~ \pm \frac 12 ~\frac\pi{\pi-\gamma} ~\left(\theta_\pm
 \mp S \gam \right)^2 ~\mp~
\frac{\pi^2}{12} ~.
\ee

%{{{ excited figure

\begin{figure}[b]
\setlength{\unitlength}{\figunit}
\noindent
\begin{picture}(150,28)
%\begin{picture}(150,55)

%distribution of k-string

%\put(97,40.1){\vector(1,0){5}}
%\put(99,42.2){$z_k$}

%\thicklines
%\put(55,40){\line(1,0){45}}

%\put(59,37){\line(0,1){6}}
%\put(51,32){$z_k(-\infty)$}

%\put(65,38.6){$\overbrace{{\bf\rm x}}^{k-{\rm string~ at~} z_k = 0 }$}

%\put(94,37){\line(0,1){6}}
%\put(87,32){$z_k(\infty)$}

%distribution of 1-strings

\thinlines
\put(147,15.1){\vector(1,0){5}}
\put(149,17.2){$z_1$}

\thicklines
\put(5,15){\line(1,0){145}}

\put(9,12){\line(0,1){6}}
\put(1,7){$z_1(-\infty)$}

\put(13,15){\circle{2}}
\put(18,15){\circle{2}}
\put(23,15){\circle{2}}
\put(28,15){\circle{2}}
\put(33,15){\circle{2}}

\put(11.5,13){$\overbrace{\phantom{12345678}}^{h^- ~{\rm holes}}$}

\put(34.9,13){$\underbrace{\phantom{123456789012345678901234567890}}_{{\rm
bulk~of~} N-S-k n_k {\rm ~real~roots}}$}
\put(36.4,13.6){{\bf x}}
\put(41.4,13.6){{\bf x}}
\put(46.4,13.6){{\bf x}}
\put(51.4,13.6){{\bf x}}
\put(56.4,13.6){{\bf x}}
\put(61.4,13.6){{\bf x}}
\put(66.4,13.6){{\bf x}}
\put(71.4,13.6){{\bf x}}
\put(76.4,13.6){{\bf x}}
\put(81.4,13.6){{\bf x}}
\put(86.4,13.6){{\bf x}}
\put(91.4,13.6){{\bf x}}
\put(96.4,13.6){{\bf x}}
\put(101.4,13.6){{\bf x}}
\put(106.4,13.6){{\bf x}}
\put(111.4,13.6){{\bf x}}
\put(116.4,13.6){{\bf x}}

\put(123,15){\circle{2}}
\put(128,15){\circle{2}}
\put(133,15){\circle{2}}
\put(138,15){\circle{2}}
\put(143,15){\circle{2}}

%\put(114.5,13){$\overbrace{\phantom{12345678}}^{h^+ = 2k-2 - h^-
\put(121.5,13){$\overbrace{\phantom{12345678}}^{h^+
~{\rm ~ holes}}$}

\put(144,12){\line(0,1){6}}
\put(137,7){$z_1(\infty)$}

\end{picture}

\caption{{\it The distribution of real roots and holes in a  state
with one $k$-string.}}

\smallskip

\end{figure}
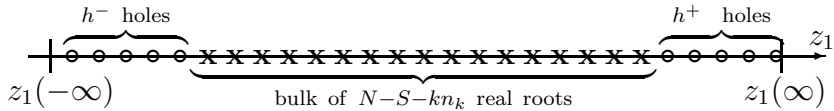

%}}}

Combining the contributions of the various terms in Eq. (\ref{calE}),
the finite size energy of the state described above evaluates to
\bea
 E_{{\rm FSC}} = \frac{1}{2 N} \Biggl\{
 \frac 1{\pi-\gam} \biggl[ \theta  + \pi \Delta h
                            - \pi(k-1) \Delta n_k \biggr]^2
 ~+~ (\pi-\gam) S^2 ~-~  \frac\pi 6  ~ ~ ~ ~ ~ ~ ~ ~ ~ ~ ~ ~ ~ ~
\cr
 +~ 2\pi \Biggl( k \mm \Delta h \mm \Delta n_k  + S n_k (k-1)
 - n^+_k n^-_k (1+ 2 k - 2 k^2)
 + \sum_{j=1}^{n_k^+} I_{j}^{(k)+}
 - \sum_{j=1}^{n_k^-} I_{j}^{(k)-} \Biggr)
\Biggr\}
\mabel{fscenergy}
\eea
 where the antisymmetry of the $k$-string distribution is $\Delta n_k
= n_k^+ - n_k^-$, and the antisymmetry of the hole distribution is
 \be
 \Delta h ~=~  \frac 12 \left( h^+ - h^-
 +\left\lfloor \frac 12-\frac\theta\pi+S\frac\gam\pi \right\rfloor
 +\left\lfloor \frac 12-\frac\theta\pi-S\frac\gam\pi \right\rfloor
 \right) ~ . \mabel{deltah}
 \ee

With the distribution of real roots as described, it is an easy
task to compute the momentum (\ref{moment-mal}). This turns out to
be
\be
P= \pi \left( N-S-k n_k +1 - \Delta h \right) ~\mod 2\pi
 ~+~ \frac{\pi}{N}  \biggl(
 k n_k \mm \Delta h ~+~ S \biggl(\frac\theta\pi +  \mm\Delta h \biggr)
 ~+~ \sum_{j=1}^{n_k} I_j^{(k)}
 \biggr) \mabel{finmom}
 \ee

The twisted ground state is the state with no strings nor holes. For
zero external magnetic field, it lies in the sector $S=0$, and has
vanishing momentum. The ground state energy is
 \be
 E^{(0,\theta)}_{{\rm FSC}} = \frac1{2 N} \left(
\frac{\theta^2  }{\pi-\gam} ~-~ \frac \pi{6} \right)
 \mabel{groundE}
 \ee
 The central charge is $c=1$, as the well-known formula of Refs.
\cite{blocani,affleck} connecting the ground state energy with the
central charge gets corrected in the presence of non-trivial boundary
conditions. The central charge is a bulk property, which cannot be
changed by boundary effects.

For the subsets of parameter space investigated earlier, the computed
excited state energies and momenta agree with results in the
literature. For the pure hole states (no complex excitations), they
coincide with the ones in Ref. \cite{eckham}. Regarding states with
complex excitations, the two-string result agrees with Ref.
\cite{kluwezi}. For generic strings, our results agree with the ones
of Ref. \cite{karowski}, for the discrete set of couplings ($\gam =
\pi/\nu, ~ \nu= 3,4, \ldots$) and twists ($\theta = \gam$) treated
there. When comparing with the literature, it should be noted that our
definition of the Hamiltonian by differentiation (\ref{ene-ene})
rescales the Fermi-velocity to 1.

It is also interesting to compare to results obtained by the
bosonisation method. From e.g. \cite{eggafl} one can find the
excitation spectrum for the untwisted chain. The result for a twisted
chain follows from observing that the twist boundary conditions on the
chain modify the boundary conditions of the dual boson only. This
leads to an energy spectrum which is exactly of the form
(\ref{fscenergy}),
\be
 E_{{\rm bosonisation}} = \frac{v_s}{2 N} \left(
 \frac{(m+\theta)^2}{2 \pi R^2}
+ 2 \pi R^2 S^2 + 2\pi \sum_{n=1}^\infty n(m_n^L + m_n^R) \right) ~ .
\ee
 Here $R=\sqrt{\frac{\pi-\gam}{2\pi}}$ is the compactification radius
of the boson, $m$ is the quantized momentum of the boson zero-mode,
and $m_n^{L,R}$ are occupation numbers of the $n$:th left and right
moving oscillator modes of the boson. $S$ is the quantized dual
boson zero-mode momentum, which coincides with our $S$. Comparing to
(\ref{fscenergy}), we see that Bethe Ansatz strings and holes are
mixtures of bosonic zero and oscillatory modes.

%}}}
%{{{ The Ka\v c table in Liouville theory

\section{The Ka\v c table in Liouville theory}

As explained in the introduction, imaginary coupling
Liouville theory can be described by a collection of
twisted XXZ chains, if $\gam/\pi = \mu/(\nu+1)$ is rational
\cite{fadtir}.\footnote{More exactly, in \cite{fadtir} the
Liouville equivalence was proved only for $\pi/\gam = \nu
+1$, $\nu=2,3,\ \ldots$ The equivalence can be easily
extended to more general rational values.}
 The twists are quantized in terms of $\gam$;
 \[
 \theta = \kap\mm \gam, ~\kap = 0, \ldots, \nu ~ .
\]
 This collection of twisted XXZ${}_{1/2}$ chains is exactly
the one that describes the Bethe Ansatz of a root-of-unity
$U_q(sl_2)$ invariant periodic chain
\cite{karzap,grpaprra,papr}. This equivalence is not
surprising, if one keeps the connection to sine-Gordon
theory in mind. The twist-sectors in the lattice Liouville
model correspond exactly to the $\theta$-vacua of its
massive perturbation, the sine-Gordon model. These vacua
become non-degenerate in finite volume (see e.g.
\cite{fimaqura}). The sine-Gordon model has an $U_q(sl_2)$
symmetry \cite{smirnov,berlecl}, where $q$ is related to
the coupling constant $\beta^2 \equiv 8(\pi - \gam)$.

On the other hand, the integrable lattice discretation of
the sine-Gordon model is an inhomogeneous XXZ${}_{1/2}$
chain with anisotropy $\gam$. The massless limit, i.e. the
imaginary coupling Liouville model, is then described by an
homogeneous XXZ${}_{1/2}$ chain with $U_q(sl_2)$ symmetry,
of the type described in Refs. \cite{karzap,grpaprra,papr}.

The consistent quantum group restriction of the Hilbert
space of these root-of-unity $U_q(sl_2)$ invariant models
comes about by restricting $\kap$ to be nonvanishing. In
sine-Gordon theory this leads to the restricted sine-Gordon
model, which flows to minimal models in the ultraviolet
\cite{smirnov,berlecl}. Here, we are investigating the
Bethe Ansatz description of this ultraviolet limit, and we
shall see that the primary operators creating string states
furnish the whole
 Ka\v c table.

The ground state of the quantum group restricted model is in the
$\kap =1$ sector, reproducing
 \be
c= 1 - 6\gam^2/\pi(\pi-\gam) ~.
% = - \frac{6 N}\pi E_{{\rm FSC}}
 \ee
 For rational $\gam/\pi$, this gives exactly the central
charges of minimal models.

Here it should be noticed that in the quantum group
invariant chain, the twist does not arise from boundary
conditions on the spatial boundaries of the chain. Rather,
different twists select different sectors in the Hilbert
space transforming in a specific way under the global
$\ZZ$-symmetry of the theory. In a bosonized language this
symmetry is $\phi \to \phi + 2 n \pi, ~ n\in\ZZ$. This is a
symmetry of all theories with the potential a function of
$\e{\i\phi}$, including the sine-Gordon and imaginary
Liouville ones. Thus the situation for a restricted quantum
group invariant chain (i.e. a q-restricted Liouville chain)
differs from the XXZ${}_{1/2}$ chain with twisted spatial
boundary conditions discussed in the previous Section. The
equation of Refs. \cite{blocani,affleck}, connecting the
ground state energy to the central charge, does not get
boundary corrections, and the central charge is as above.

Conformal primary and secondary operators create excited
states from the vacuum. Following Cardy \cite{cardy}, the
critical indices of the operators creating excited states
are given by the finite size energies and momenta.

We are interested in primary states with equal holomorphic
and anti-holomorphic confomal weights, i.e. with vanishing
finite size correction to the momentum (conformal spin).
Inspired by Ref. \cite{karowski}, we look for these among
the excited states with one higher string ($n_k=1$). From
equation (\ref{finmom}) we see that a sufficient condition
for the vanishing of the conformal spin is $S=0$ and
 \be
 k \mm\Delta h + I^{(k)} = 0    ~. \mabel{nospin}
\ee
 The corresponding energies are
\bea
 E_{{\rm FSC}} = \frac{1}{2 N} \Biggl\{
 \frac 1{\pi-\gam} \biggl[ \kap \gam - \pi I^{(k)\pm}/k
\mp  \pi (k-1)
\biggr]^2  - \frac\pi 6 \Biggr\} ~,
\eea
 where the sign refers to the position of the string close to $\lam =
\pm\infty$. The harmonic oscillator pieces in the energies vanish for
all of these states.

For generic values of $\theta$ and $\gam<\pi/2$, the $k$-string may be
close to $\infty$ or $-\infty$, depending on the values of $k,\gam$
and $\theta$. Not all string legths are allowed for all
$\gam$ and $\theta$, nor all combinations of different hole and string
positions. Apart from the Takahashi-Suzuki conditions \cite{taksuz}
that restrict allowed string lengths, one has to treat the Bethe
Ansatz equation for $z_k$ to find the allowed values of $I_k$.

For our purposes, it suffices to analyze the asymptotics of
$z_k(\zeta)$ for the string position $\zeta$ close to
$\pm\infty$, which gives the possible range of quantum
numbers $I^{(k)}$. We get different branches of primary
states solving the condition (\ref{nospin}).

The first branch, for which $I^{(k)} =0$, exists
when
\be
 k + \kap < \frac\pi\gam  ~ .\mabel{karR}
\ee
 For states with hole distributions symmetric up to the
twist effects of Equation (\ref{nospin}), this
$k$-string is close to $-\infty$.

Using the formula
$E_{{\rm exited}} - E_o = \frac{\pi}N (\Delta +\bar\Delta)$
 %$P_{{\rm exited}} - P_o = \frac{\pi}N (\Delta -\bar\Delta)$
 of Ref. \cite{cardy}, we get the conformal weights of
these primary states:
\be
 \Delta = \bar\Delta
 = \frac{\bigl( \pi(k-1) - \kap\gam \bigr)^2 -\gam^2}{4\pi(\pi-\gam)}
 ~.
\ee
 For unitary minimal models ($\gam = \frac\pi{\nu+1}$),
these states consisting of one $k$-string with $I^{(k)} =
0$ and a symmetric distribution of $2k-2$ holes, furnish
the whole
 Ka\v c table. In the q-restricted case, the remnant $\kap$
gets the values $\kap = 1, \ldots , \nu$. The
Takahashi-Suzuki condition \cite{taksuz} on the string
length allows all strings up to $k=\nu$. For
an unitary model, Condition (\ref{karR}) restricts this
further to $k < \nu+1-\kap$. Thus the integers $\kap$ and
$k$ give exactly the two integer labels of the full
 Ka\v c table, with the apropriate ranges.

For non-unitary minimal models, these states yield only a
part of the table. Some primaries are left out due to the
technical restriction adopted in this paper, $k\gamma<\pi$.

%}}}
%{{{ Conclusions

\section{Conclusions}

We have computed the finite size corrections to the
energies and momenta of excited states in XXZ${}_{1/2}$
chains periodic up to a twist, for generic anisotropy $\gam
< \pi/2$ and twist $\theta$. The excited states were
described as holes in the sea of real Bethe Ansatz
rapidities, as well as complex rapidities collected in
(positive parity) strings. We restricted outselves to the
case where there are strings of only one spcified length
$k$ in addition to the real rapidities.

For extracting the scaling information, we used the
Destri-de Vega approach to treat Bethe Ansatz equations,
generalizing it to cope with string-like excited states.

Apart from completing the picture of excitation energies to
be found from the literature, to arbitrary coupling, twist
and string length (up to $k< \pi/\gam$), our  result
(\ref{fscenergy},\ref{finmom}) is interesting due to its
special dependence of the ``field theoretic'' quantum
numbers $h^\pm$. This we consider the major novel feature
of our finite size solution.

One of the original motivations to develop the DdV
formalism was the possibility to derive thermodynamic Bethe
Ansatz (TBA) equations directly from the albegraic Bethe
Ansatz \cite{desdeve}. The field theoretic TBA degrees of
freedom correspond to holes and strings, which are
excitations above the vacuum of interacting magnons. The
holes correspond to right and left moving kinks, and the
strings correspond in this case to weakly bound kinks, with
binding energy of the order $\sim 1/N$.

As can be seen from Equation (\ref{deltah}), the twist
dependence of the energy for a fixed field-theoretical
configuration is not innocent. Due to the integer part
cuts, the energies and momenta have discontinuities at some
specific values of the twist. From the point of view of the
original magnon degrees of freedom $\lambda_j^{(l)}$ of the
Bethe Ansatz equations (\ref{IBAE}), however, an adiabatic
excursion in $\theta$ is possible. The $\lambda_j^{(l)}$:s
are the building blocks of the underlying interacting
vacuum, and the energies and momenta are continuous if the
$\lambda_j^{(l)}$:s move continuously.

During such an adiabatic excursion, the number of left
and/or right holes change when $\theta$ moves over a value
giving a jump in the integer part expressions of
(\ref{deltah}). Thus at these values of $\theta$ the vacuum
absorbs a left kink state and emits a right one, or vice
versa. It should be stressed that this spectral flow is not
just a trivial consequence of the choice of branch in the
counting functions. Rather, it encodes drastic changes in
the structure of the Hilbert space for different values of
$\theta$. For example, the degeneracies of secondary states
built by driving some of the holes into the bulk change at
the jump points.

Adiabatic excursions of this type have been investigated in
the context of a XXZ chain threaded by a magnetic flux
\cite{suthshast,yufow,roecksuth}. For this, the twist is
given the physical interpretation as a threading magnetic
flux. For full comparison with the numeric excited state
results of \cite{yufow,roecksuth}, negative parity strings
have to be encorporated into our treatment, as these play a
central role for larger values of $\gam$.

Finally, we showed how in the q-restricted Hilbert space of
the $U_q(sl_2)$ invariant periodic chain, which is
equivalent to the imaginary coupling Liouville model and
the UV-limit of sine-Gordon, the central charge decreases
from $c=1$ to those of minimal models. Excited states with
one higher string and a minimal amount of holes, give rise
to all the primary states in the case of unitary minimal
models.

To conclude, we want to comment that the string picture
should not be necessary for deriving the results of this
paper. In Ref. \cite{ddv2}, wide and close pairs were
used to describe complex excitations. This approach
could certainly be generalized to twisted boundary
conditions in the XXZ chain. The strings, however,
provide a good tool which singles out very specific
combinations of wide and close pairs that correspond to
primary states in the critical theory.

%}}}

%{{{ Acknowledgement

\section*{Acknowledgements}

We thank L.D. Faddeev, H.-P. Eckle, A. Luther, C. Montonen,
A. Neveu, H.J. de Vega and Al.B. Zamolodchikov for
conversations on the subject of this paper. T.N.
acknowledges the hospitality of NORDITA during a visit
where part of this work was done.

%}}}

%{{{ References

%\pagebreak[4]

%}}}

\end{document}